\documentclass{webofc}
\usepackage[varg]{txfonts}   
\usepackage{amssymb, amsmath}
\usepackage{hyperref}
\usepackage{textcomp}
\usepackage{datatool}

\DTLsetseparator{   }
\DTLloaddb{NGC2146}{NGC2146variables.txt}
\DTLloaddb{NGC2976}{NGC2976variables.txt}

\begin{document}
\begin{filecontents*}{NGC2146variables.txt}
variable	value
COcontr_mean	16
COcontr_std	7
COcontr_global	18
mass_mean	2.65
mass_std	2.33
temp_mean	29.61
temp_std	2.78
beta_mean	1.90
beta_std	0.11
thfr_mean	8.82
thfr_std	6.92
temp_min	22.7
temp_min_er	3.2
temp_min_el	2.4
temp_max	35.7
temp_max_er	1.6
temp_max_el	1.5
mass_max	10.01
mass_max_er	0.41
mass_max_el	0.42
beta_max	2.13
beta_max_er	0.20
beta_max_el	0.24
beta_min	1.57
beta_min_er	0.34
beta_min_el	0.30
calibfac_1	1.15
calibfac_2	36.25
ecalibfac_1	4.09
ecalibfac_2	110.93
calibpow_1	1.31
calibpow_2	0.99
ecalibpow_1	0.15
ecalibpow_2	0.13

T	29.35
T_le	1.06
T_re	1.21
M	5.22
M_le	0.267
M_re	0.254
beta	2.00
beta_le	0.10
beta_re	0.09
A1	0.142
A1_le	0.021
A1_re	0.019
A2	1.341
A2_le	0.071
A2_re	0.073
alpha	1.00
alpha_le	0.08
alpha_re	0.09
relres_1	-6
relres_2	3
\end{filecontents*}

\begin{filecontents*}{NGC2976variables.txt}
variable	value
COcontr_mean	5
COcontr_std	2
COcontr_global	5
mass_mean	2.00
mass_std	0.75
temp_mean	27.68
temp_std	1.86
beta_mean	1.38
beta_std	0.13
thfr_mean	3.90
thfr_std	2.93
temp_min	23.5
temp_min_er	4.8
temp_min_el	3.4
temp_max	33.5
temp_max_er	4.6
temp_max_el	3.6
mass_max	4.46
mass_max_er	0.38
mass_max_el	0.38
beta_min	1.04
beta_min_er	0.34
beta_min_el	0.28
beta_max	1.74
beta_max_er	0.35
beta_max_el	0.28
calibfac_1	22.26
calibfac_2	4.23
ecalibfac_1	22.26
ecalibfac_2	6.54
calibpow_1	0.61
calibpow_2	0.82
ecalibpow_1	0.05
ecalibpow_2	0.07
H2mass	1.02
HImass	6.37
gasmass	1.65
DGRmean	0.004
DGRstd	0.001
globalDGR	0.003
dustHI_a	0.49
dustHI_ae	0.03
dustHI_b	3.60
dustHI_be	0.10
dustHI_r	0.721
dustHI_re	0.044
dustH2_a	1.03
dustH2_ae	0.05
dustH2_b	2.08
dustH2_be	0.17
dustH2_r	0.791
dustH2_re	0.044
HIH2_a	1.18
HIH2_ae	0.10
HIH2_b	-0.71
HIH2_be	0.50
HIH2_r	0.617
HIH2_re	0.044

T	26.11
T_le	0.88
T_re	0.92
M	1.07
M_le	0.059
M_re	0.057
beta	1.44
beta_le	0.08
beta_re	0.07
A1	0.019
A1_le	0.011
A1_re	0.007
A2	0.080
A2_le	0.013
A2_re	0.013
alpha	1.16
alpha_le	0.37
alpha_re	0.37
relres_1	-3
relres_2	0


\end{filecontents*}

    \title{Constraining Millimeter Dust Emission in Nearby Galaxies with NIKA2: the case of NGC2146 and NGC2976}

    \author{\lastname{G.~Ejlali}\inst{\ref{IPM}}\fnsep\thanks{\email{gejlali@ipm.ir}}
              \and  R.~Adam \inst{\ref{OCA}}
              \and  P.~Ade \inst{\ref{Cardiff}}
              \and  H.~Ajeddig \inst{\ref{CEA}}
              \and  P.~Andr\'e \inst{\ref{CEA}}
              \and  E.~Artis \inst{\ref{LPSC},\ref{Garching}}
              \and  H.~Aussel \inst{\ref{CEA}}
              \and  M.~Baes  \inst{\ref{Belguim}}
              \and  A.~Beelen \inst{\ref{LAM}}
              \and  A.~Beno\^it \inst{\ref{Neel}}
              \and  S.~Berta \inst{\ref{IRAMF}}
              \and  L.~Bing \inst{\ref{LAM}}
              \and  O.~Bourrion \inst{\ref{LPSC}}
              \and  M.~Calvo \inst{\ref{Neel}}
              \and  A.~Catalano \inst{\ref{LPSC}}
              \and  M.~De~Petris \inst{\ref{Roma}}
              \and  F.-X.~D\'esert \inst{\ref{IPAG}}
              \and  S.~Doyle \inst{\ref{Cardiff}}
              \and  E.~F.~C.~Driessen \inst{\ref{IRAMF}}
              \and  F.~Galliano \inst{\ref{CEA}}
              \and  A.~Gomez \inst{\ref{CAB}} 
              \and  J.~Goupy \inst{\ref{Neel}}
              \and  A.~P.~Jones \inst{\ref{IAS}}
              \and  C.~Hanser \inst{\ref{LPSC}}
              \and  A.~Hughes \inst{\ref{IRAP}}
              \and  S.~Katsioli \inst{\ref{Athens_obs},\ref{Athens_univ}}
              \and  F.~K\'eruzor\'e \inst{\ref{Argonne}}
              \and  C.~Kramer \inst{\ref{IRAMF}}
              \and  B.~Ladjelate \inst{\ref{IRAME}} 
              \and  G.~Lagache \inst{\ref{LAM}}
              \and  S.~Leclercq \inst{\ref{IRAMF}}
              \and  J.-F.~Lestrade \inst{\ref{LERMA}}
              \and  J.~F.~Mac\'ias-P\'erez \inst{\ref{LPSC}}
              \and  S.~C.~Madden \inst{\ref{CEA}}
              \and  A.~Maury \inst{\ref{CEA}}
              \and  P.~Mauskopf \inst{\ref{Cardiff},\ref{Arizona}}
              \and  F.~Mayet \inst{\ref{LPSC}}
              \and  A.~Monfardini \inst{\ref{Neel}}
              \and  A.~Moyer-Anin \inst{\ref{LPSC}}
              \and  M.~Mu\~noz-Echeverr\'ia \inst{\ref{LPSC}}
              \and  A.~Nersesian \inst{\ref{Belguim}}
              \and  L.~Pantoni \inst{\ref{CEA}}
              \and  D.~Paradis \inst{\ref{IRAP}}
              \and  L.~Perotto \inst{\ref{LPSC}}
              \and  G.~Pisano \inst{\ref{Roma}}
              \and  N.~Ponthieu \inst{\ref{IPAG}}
              \and  V.~Rev\'eret \inst{\ref{CEA}}
              \and  A.~J.~Rigby \inst{\ref{Leeds}}
              \and  A.~Ritacco \inst{\ref{ENS}, \ref{INAF}}
              \and  C.~Romero \inst{\ref{Pennsylvanie}}
              \and  H.~Roussel \inst{\ref{IAP}}
              \and  F.~Ruppin \inst{\ref{IP2I}}
              \and  K.~Schuster \inst{\ref{IRAMF}}
              \and  A.~Sievers \inst{\ref{IRAME}}
              \and  M.~W.~S.~L.~Smith \inst{\ref{Cardiff}}
              \and  F.~S.~Tabatabaei \inst{\ref{IPM}}
              \and  J.~Tedros \inst{\ref{IRAME}}
              \and  C.~Tucker \inst{\ref{Cardiff}}
              \and  E.~M.~Xilouris \inst{\ref{Athens_obs}}
              \and  R.~Zylka \inst{\ref{IRAMF}}
}
\institute{
    Institute for Research in Fundamental Sciences (IPM), School of Astronomy, Tehran, Iran
    \label{IPM}
    \and
    Universit\'e C\^ote d'Azur, Observatoire de la C\^ote d'Azur, CNRS, Laboratoire Lagrange, France 
    \label{OCA}
    \and
    School of Physics and Astronomy, Cardiff University, CF24 3AA, UK
    \label{Cardiff}
    \and
    Universit\'e Paris-Saclay, Université Paris Cité, CEA, CNRS, AIM, 91191, Gif-sur-Yvette, France
    \label{CEA}
    \and
    Universit\'e Grenoble Alpes, CNRS, Grenoble INP, LPSC-IN2P3, 38000 Grenoble, France
    \label{LPSC}
    \and	
    Max Planck Institute for Extraterrestrial Physics, 85748 Garching, Germany
    \label{Garching}
    \and
    Aix Marseille Univ, CNRS, CNES, LAM, Marseille, France
    \label{LAM}
    \and
    Universit\'e Grenoble Alpes, CNRS, Institut N\'eel, France
    \label{Neel}
    \and
    Institut de RadioAstronomie Millim\'etrique (IRAM), Grenoble, France
    \label{IRAMF}
    \and 
    Dipartimento di Fisica, Sapienza Universit\`a di Roma, I-00185 Roma, Italy
    \label{Roma}
    \and
    Univ. Grenoble Alpes, CNRS, IPAG, 38000 Grenoble, France
    \label{IPAG}
    \and
    Centro de Astrobiolog\'ia (CSIC-INTA), Torrej\'on de Ardoz, 28850 Madrid, Spain
    \label{CAB}
    \and
    National Observatory of Athens, IAASARS, GR-15236, Athens, Greece
    \label{Athens_obs}
    \and
    Faculty of Physics, University of Athens, GR-15784 Zografos, Athens, Greece
    \label{Athens_univ}
    \and
    High Energy Physics Division, Argonne National Laboratory, Lemont, IL 60439, USA
    \label{Argonne}
    \and  
    Instituto de Radioastronom\'ia Milim\'etrica (IRAM), Granada, Spain
    \label{IRAME}
    \and
    LERMA, Observatoire de Paris, PSL Research Univ., CNRS, Sorbonne Univ., UPMC, 75014 Paris, France  
    \label{LERMA}
    \and
    School of Earth \& Space and Department of Physics, Arizona State University, AZ 85287, USA
    \label{Arizona}
    \and
    School of Physics and Astronomy, University of Leeds, Leeds LS2 9JT, UK
    \label{Leeds}
    \and
    INAF-Osservatorio Astronomico di Cagliari, 09047 Selargius, Italy
    \label{INAF}
    \and 
    LPENS, ENS, PSL Research Univ., CNRS, Sorbonne Univ., Universit\'e de Paris, 75005 Paris, France 
    \label{ENS}
    \and  
    Department of Physics and Astronomy, University of Pennsylvania, PA 19104, USA
    \label{Pennsylvanie}
    \and
    Institut d'Astrophysique de Paris, CNRS (UMR7095), 75014 Paris, France
    \label{IAP}
    \and
    University of Lyon, UCB Lyon 1, CNRS/IN2P3, IP2I, 69622 Villeurbanne, France
    \label{IP2I}
    \and
    Sterrenkundig Observatorium Universiteit Gent, Krijgslaan 281 S9, B-9000 Gent, Belgium
    \label{Belguim}
    \and
    Institut d'Astrophysique Spatiale (IAS), CNRS, Universit\'e Paris Sud, Orsay, France
    \label{IAS}
    \and
    IRAP, Université de Toulouse, CNRS, UPS, IRAP, Toulouse Cedex 4, France
    \label{IRAP}
}

\abstract{This study presents the first millimeter continuum mapping observations of two nearby galaxies, the starburst spiral galaxy NGC2146  and the dwarf galaxy NGC2976, at 1.15\,mm and 2\,mm using the NIKA2 camera on the IRAM 30m telescope, as part of the Guaranteed Time Large Project IMEGIN. These observations provide robust resolved information about the physical properties of dust in nearby galaxies by constraining their FIR-radio SED in the millimeter domain. After subtracting the contribution from the CO line emission, the SEDs are modeled spatially using a Bayesian approach. Maps of dust mass surface density, temperature, emissivity index, and thermal radio component of the galaxies are presented, allowing for a study of the relations between the dust properties and star formation activity (using observations at 24$\mu$m as a tracer). We report that dust temperature is correlated with star formation rate in both galaxies. The effect of star formation activity on dust temperature is stronger in NGC2976, an indication of the thinner interstellar medium of dwarf galaxies. Moreover, an anti-correlation trend is reported between the dust emissivity index and temperature in both galaxies.}

\maketitle

\section{Introduction}
Dust grains in galaxies are important for the formation of protostellar cores and significantly impact the heating and cooling processes of the InterStellar Medium (ISM). Modeling the Spectral Energy Distribution (SED) of dust allows us to infer the physical properties of the grains. Dust emits across a range of temperatures, with the warmer component emitting in the Mid-InfraRed (MIR) and colder dust emitting in the Far-InfraRed (FIR). The sub-millimeter/millimeter waveband is crucial for detecting this cold dust component and estimating the total dust mass. Telescopes such as IRAS, ISO, \textit{Spitzer}, and \textit{Herschel} have allowed us to study dust emission up to 500\,$\mu$m, but studying longer millimeter wavelengths is crucial for modeling the mass and temperature of cold dust and constraining the radio component \cite{Galliano+2018}. \\
The New IRAM KIDs Array (NIKA2) on the IRAM 30m telescope brings the unique opportunity to map full galaxies for the first time at 1.15\,mm and 2\,mm. \cite{NIKA2-performance}. Within the Guaranteed Time Large Project IMEGIN (Interpreting Millimeter Emission of Galaxies with IRAM and NIKA2, PI: S.~Madden), a sample of 22 nearby galaxies with varying ranges of mass, morphological types, star formation rate (SFR), and ISM properties have been observed. This paper presents observations of two IMEGIN galaxies of widely different physical properties: NGC2146 a starburst spiral galaxy (D=3.5\,Mpc), and NGC2976 a peculiar dwarf galaxy (D=17.2\,Mpc) (Ejlali \textit{et al.} in prep.).\\

\begin{figure}
\centering
	\includegraphics[width=0.4\textwidth]{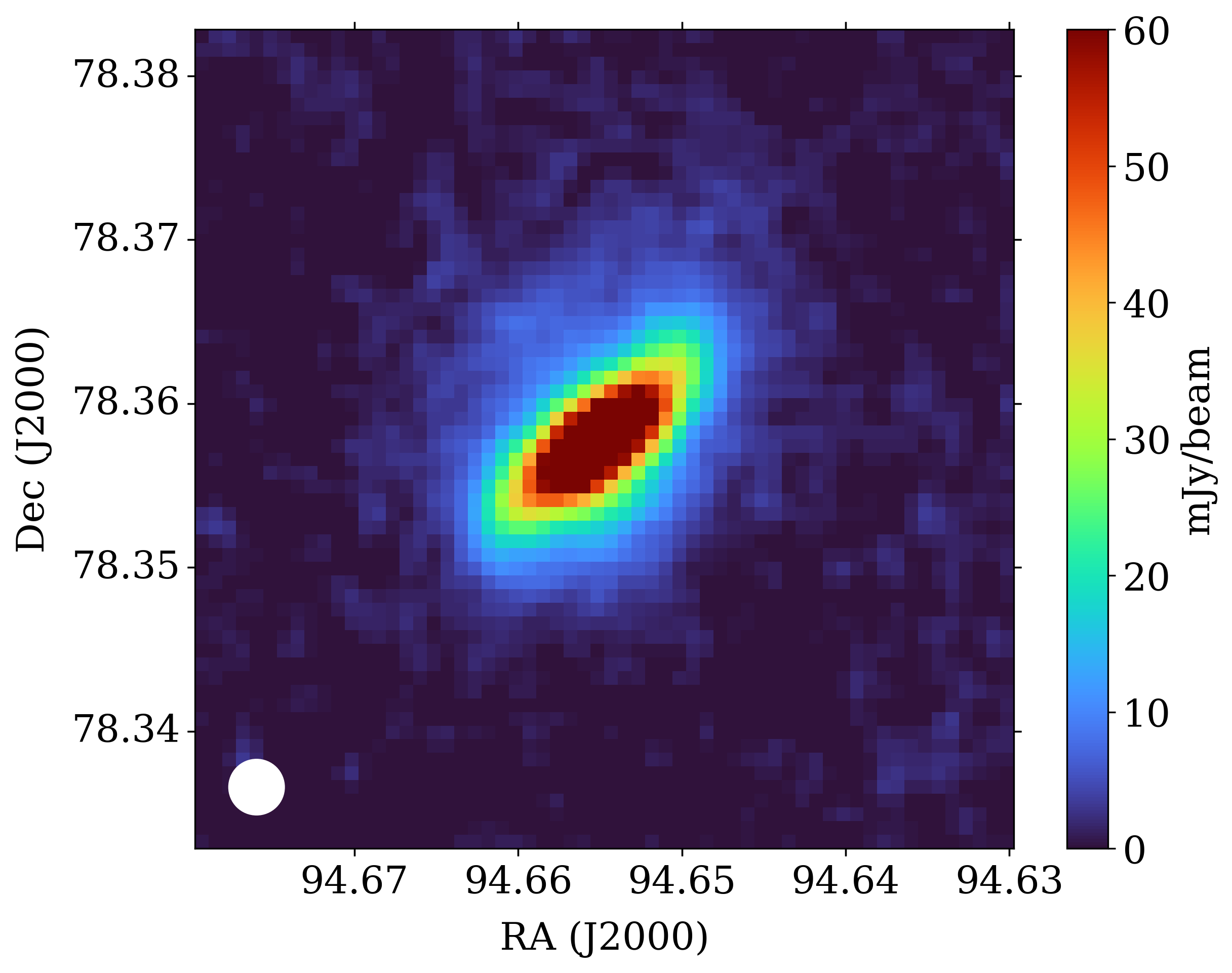}
	\includegraphics[width=0.4\textwidth]{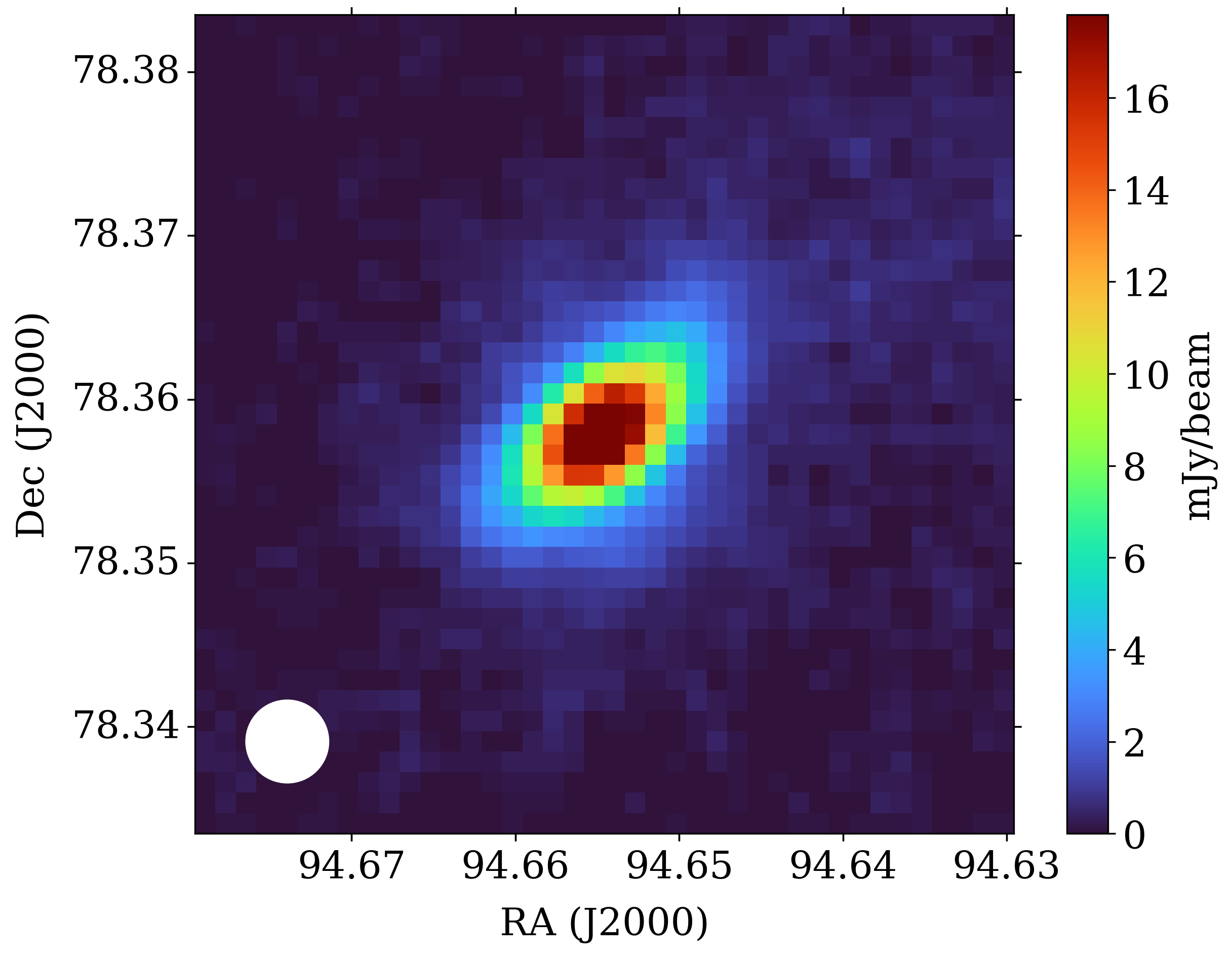}
	\includegraphics[width=0.4\textwidth]{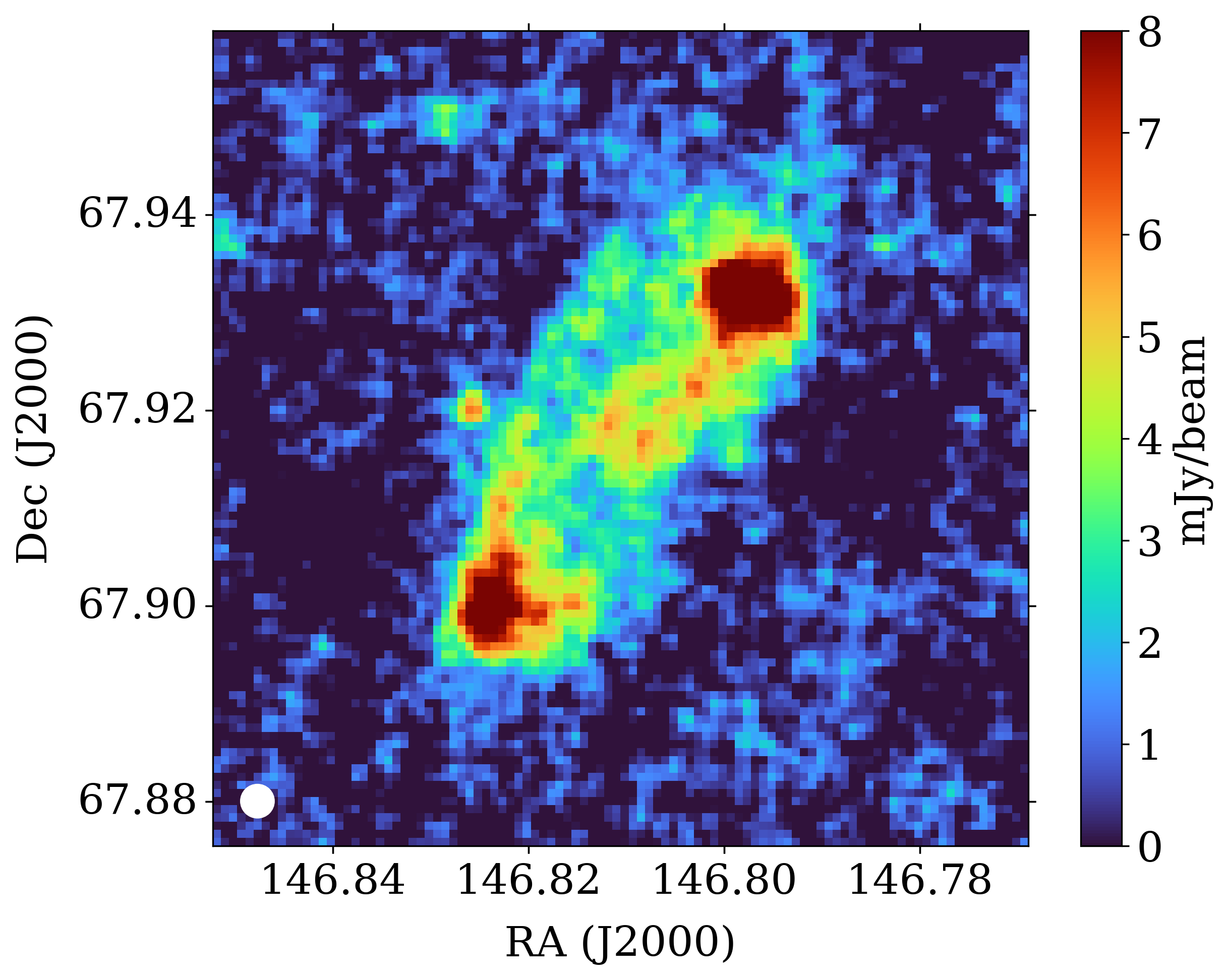}
	\includegraphics[width=0.4\textwidth]{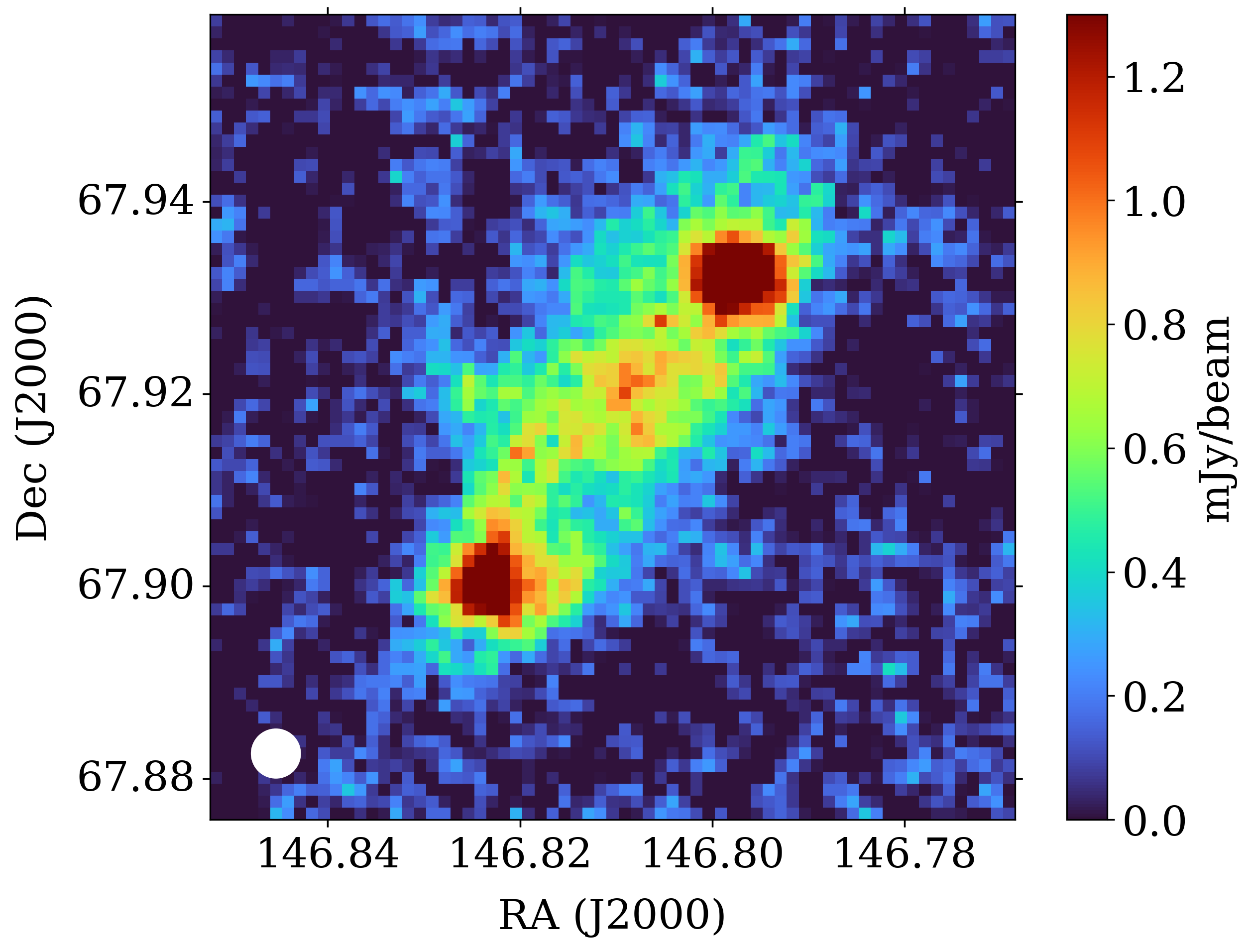}
\caption{Observed NIKA2 maps of NGC2146 (\textit{top}) and NGC2976 (\textit{bottom}) at  1.15\,mm (\textit{left}, resolution 12$^{\prime\prime}$) and 2\,mm (\textit{right}, resolution 18$^{\prime\prime}$). The rms noise level for NGC2146 are 0.9~(1.15\,mm) and 0.24~(2\,mm) mJy/beam and for NGC2976 is 0.8~(1.15\,mm) and 0.23~(2\,mm) mJy/beam.}
\label{observedmaps}
\end{figure}

\section{Data}
We observed NGC2146 and NGC2976 with the NIKA2 camera for 5 and 5.5 hours, respectively. The data were reduced using \textit{Scanam\_NIKA} pipeline \cite{Roussel+2020} and are shown in Fig.~\ref{observedmaps}.\\
The complementary data used in this work to model the spatial SEDs include \textit{Spitzer} MIPS observations at 24\,$\mu m$ \cite{Clark+2018}, \textit{Herschel} PACS and SPIRE observations at 70, 100, 160, and 250\,$\mu$m \cite{Kennicutt+2011}, Planck observations at 1.38\,mm \cite{Clark+2018}, Radio Continuum (RC) observations with WSRT at 18\,cm and 21\,cm \cite{Braun+2007}, Effelsberg 100m telescope observations at 6.2\,cm \cite{Tabatabaei+2017}, CO(2-1) data from the HERACLES survey \cite{Leroy+2009}. All the maps are pre-processed to have the same resolution (18$^{\prime\prime}$) and geometry (pixel size 6$^{\prime\prime}$). A 6$^{\prime\prime}$ pixel size corresponds to a physical size of about $100pc$ and $500pc$ in NGC2976 and NGC2146, respectively. \\
As the NIKA2 bandpass (\cite{NIKA2-performance}) encompasses CO(2-1) line, part of the observed flux  at 1.15\,mm is due to line emission~\cite{Drabek+2012}. The mean contribution of the  CO(2-1) line emission to the observed NIKA2 continuum emission at 1.15\,mm (within pixels above 3$\sigma$ limit) is $(\DTLfetch{NGC2146}{variable}{COcontr_mean}{value}\pm \DTLfetch{NGC2146}{variable}{COcontr_std}{value})\%$ (error is the standard deviation) in NGC2146 and $(\DTLfetch{NGC2976}{variable}{COcontr_mean}{value} \pm \DTLfetch{NGC2976}{variable}{COcontr_std}{value})\%$ for NGC2976. Before including the NIKA2 1.15\,mm data point in the SED modeling process, we subtract the contamination by CO. 

\section{Model}
Continuum emission at 1.15\,mm and 2\,mm consists of radio continuum emission $S^{\rm RC}_{\nu}$ and thermal emission from dust $S^{\rm dust}_{\nu}$. To describe the radio continuum emission, we employ two power-laws defined as $S^{\rm RC}_{\nu}=A_{1}\nu^{-0.1}+A_{2}\nu^{-\alpha_{syn}}$, in which $A_1$ and $A_2$ are free parameters quantifying contributions from thermal free-free and nonthermal synchrotron components. $\alpha_{syn}$ is the synchrotron spectral index, equal to 0.71 and 1.13 for NGC2146 and NGC2976, respectively \cite{Tabatabaei+2013b, Tabatabaei+2017}. We model the thermal emission of dust using a Modified Black-Body (MBB) model defined as $S^{\rm dust}_{\nu}=\kappa_{0}\left(\frac{\nu}{\nu_{0}}\right)^{\beta} \left(\frac{M}{D^{2}}\right) B_{\nu}(T)$ with dust temperature $T_{\rm dust}$, dust mass $M_{\rm dust}$, and dust emissivity index $\beta$ as free parameters [$\kappa_{0}(250\,\rm{GHz})=0.4\,{\rm m}^{2}/{\rm kg}$]~\cite{Hunt+2015}. We find the best-fit values for the five free parameters ($M_{\rm{dust}}$, $T_{\rm{dust}}$, $\beta$, $A_{1}$, $A_{2}$) with the Bayesian approach. We use the MCMC method via Python package \textit{emcee} and fit the model to the observations \cite{emcee}.\\

\section{Results}
\begin{figure}
\centering
\includegraphics[width=0.8\textwidth]{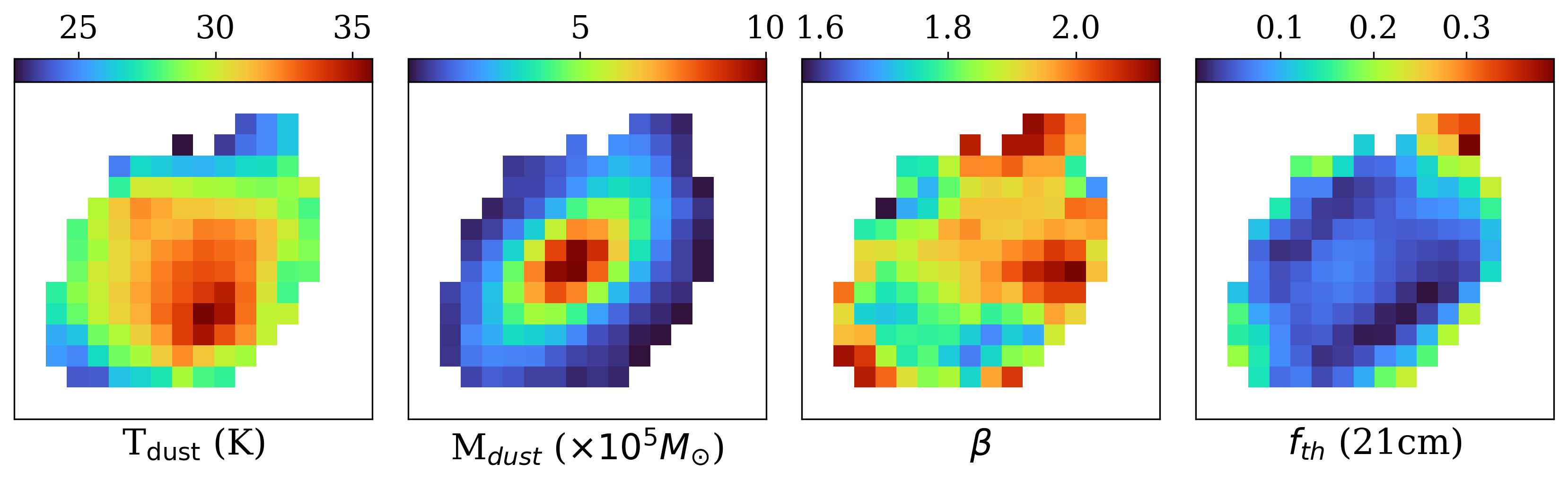}
\includegraphics[width=0.8\textwidth]{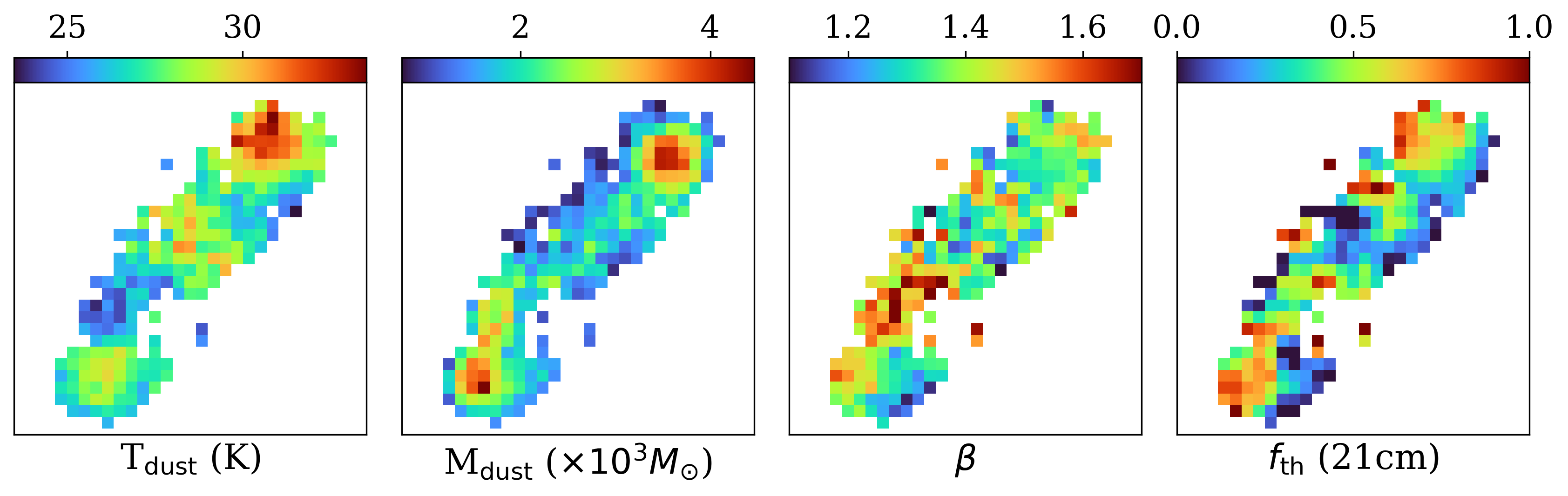}
\caption{Maps of dust temperature ($T_{\rm{dust}}$), mass ($M_{\rm{dust}}$) and emissiviry index ($\beta$) and thermal fraction ($f_{\rm th}$) at 21\,cm for NGC2146 (\textit{top}) and NGC2976 (\textit{bottom}). They are determined with the Bayesian approach and using the MCMC method, in each pixel with value larger than $3\sigma$ rms of all used maps.}
\label{parametermaps}
\end{figure}

Fig.~\ref{parametermaps} shows the best-fit values of four of the free parameters of our model across the two galaxies, namely $T_{\rm{dust}}$, $M_{\rm{dust}}$, $\beta$, and thermal free-free fraction at 21\,cm defined as $f_{\rm{th}}\,(21\,cm)=A_{1}\nu_{21\,cm}^{-0.1}/S^{\rm{RC}}_{21\,cm}$.
The mean value and standard deviation for both galaxies are reported in Table~\ref{paramstat}. In NGC2146, the dust mass peaks at the central region, and the inner $\sim$1\,kpc of the disk contains $\sim$40\% of the total dust mass in this galaxy. In NGC2976, dust mass peaks at two star-forming regions, which contain more than 25\% of the total dust mass. The dust mass in NGC2146 varies within a range that is three orders of magnitude larger than in NGC2976. The dust temperature varies across each galaxy, and peaks in the outer disk of NGC2146 ($\DTLfetch{NGC2146}{variable}{temp_max}{value}\pm\DTLfetch{NGC2146}{variable}{temp_max_er}{value} \,{\rm K}$) and in the northern star-forming region of NGC2976 ($\DTLfetch{NGC2976}{variable}{temp_max}{value}\pm\DTLfetch{NGC2976}{variable}{temp_max_er}{value}~{\rm K}$). Moreover, we report values of $3.7\times 10^{7}\,{\rm M}_{\odot}$ and $4.9\times10^{5}\,{\rm M}_{\odot}$ for the total dust mass (within pixels above 3$\sigma$ limit) in NGC2146 and NGC2976, respectively.  \\
Mapping dust emissivity index provides hints about grain formation and evolution, but the MBB model does not take into account the mixing of physical conditions along the line of sight. This can cause systematic underestimation of dust mass compared to more complex dust models \cite{Remy+2015}. In both galaxies, $\beta$ reaches its maximum in the outskirts of the disk. We find $\beta$ ranging from \DTLfetch{NGC2146}{variable}{beta_min}{value}$\pm$\DTLfetch{NGC2146}{variable}{beta_min_er}{value} to \DTLfetch{NGC2146}{variable}{beta_max}{value}$\pm$\DTLfetch{NGC2146}{variable}{beta_max_er}{value} in NGC2146. On the contrary, $\beta$ in NGC2976 has lower values throughout the galaxy, ranging from \DTLfetch{NGC2976}{variable}{beta_min}{value}$\pm$\DTLfetch{NGC2976}{variable}{beta_min_er}{value} to \DTLfetch{NGC2976}{variable}{beta_max}{value}$\pm$\DTLfetch{NGC2976}{variable}{beta_max_er}{value}. This is in agreement with previous studies (\cite{Remy+2015}) showing lower values of emissivity index in dwarf or low metallicity spiral galaxies. The maxima of $T_{\rm dust}$ and $\beta$ are located in different locations in both galaxies so we expect a negative correlation among these parameters; this is explored more in Sec.~\ref{sect-discussion}.

\begin{table}
\centering
\caption{Mean values of four free parameters of the pixel-by-pixel SED modeling are reported. Their standard deviation is reported as statistical error.}
\label{paramstat}
\begin{tabular}{lll}
\hline
		\hline
		SED parameter & NGC2146 & NGC2976 \\
		\hline
		$T_{\rm{dust}}\,({\rm K})$		&	$\DTLfetch{NGC2146}{variable}{temp_mean}{value} 
                            \pm  \DTLfetch{NGC2146}{variable}{temp_std}{value}$ 
                            &   $\DTLfetch{NGC2976}{variable}{temp_mean}{value} 
                            \pm  \DTLfetch{NGC2976}{variable}{temp_std}{value}$ \\
		$M_{\rm{dust}}\,({\rm M}_{\odot})$	&	$(\DTLfetch{NGC2146}{variable}{mass_mean}{value} 
                            \pm  \DTLfetch{NGC2146}{variable}{mass_std}{value}) \times 10^{5}$ 
                            &   $(\DTLfetch{NGC2976}{variable}{mass_mean}{value} 
                            \pm  \DTLfetch{NGC2976}{variable}{mass_std}{value}) \times 10^{3}$ \\
		$\beta$			&	$\DTLfetch{NGC2146}{variable}{beta_mean}{value} 
                            \pm  \DTLfetch{NGC2146}{variable}{beta_std}{value}$ 
                            &   $\DTLfetch{NGC2976}{variable}{beta_mean}{value}
                            \pm  \DTLfetch{NGC2976}{variable}{beta_std}{value}$ \\
        $f_{\rm th}$	&	$(\DTLfetch{NGC2146}{variable}{thfr_mean}{value} 
                            \pm \DTLfetch{NGC2146}{variable}{thfr_std}{value}) \times 10^{-2}$ 
                            &   $(\DTLfetch{NGC2976}{variable}{thfr_mean}{value} 
                            \pm \DTLfetch{NGC2976}{variable}{thfr_std}{value}) \times 10^{-1}$ \\
		\hline
		\noalign {\medskip}
\end{tabular}

\end{table}

\section{Discussion}\label{sect-discussion}
\begin{figure}
	\centering 
         \sidecaption
	\includegraphics[width=0.37\textwidth]{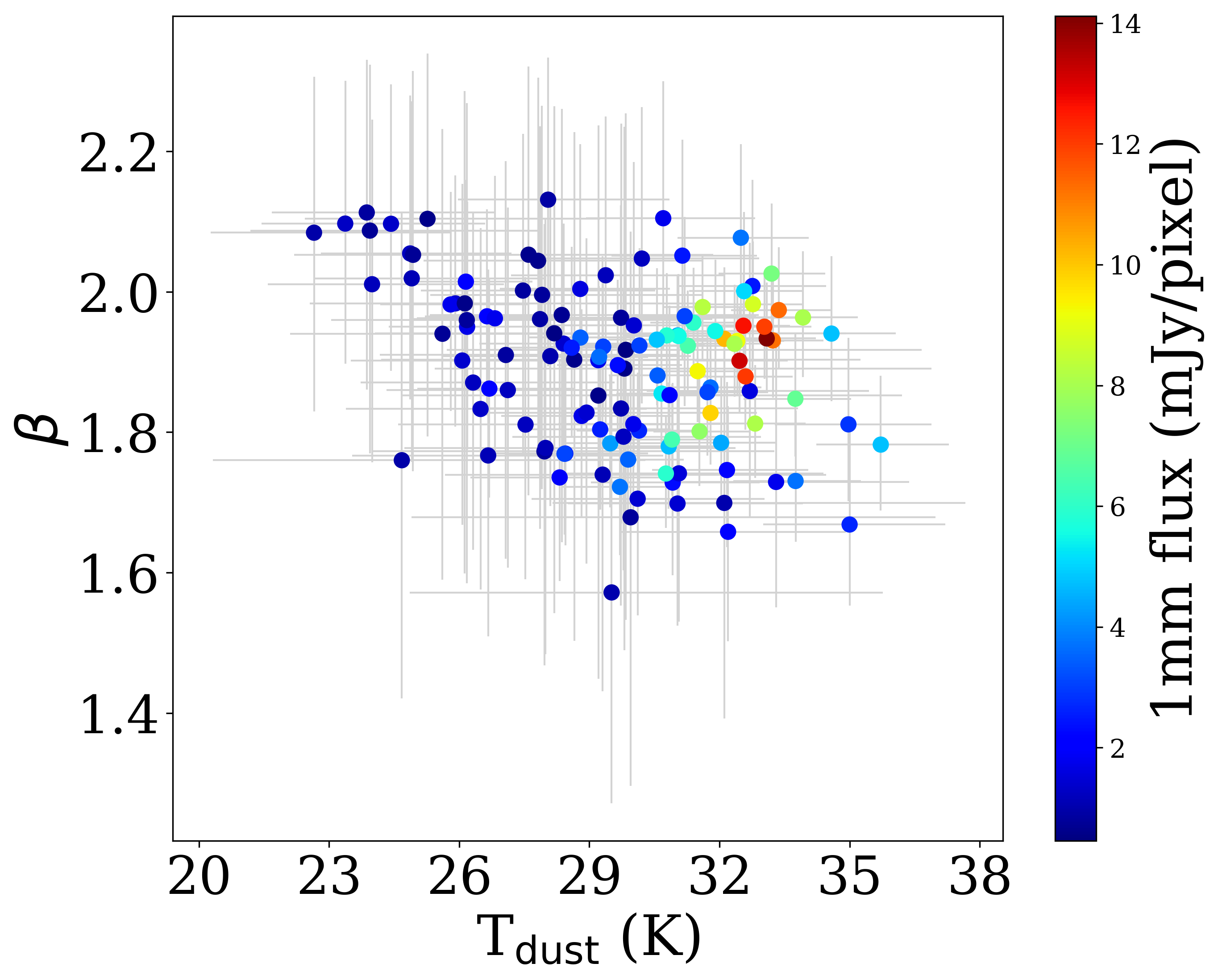}
	\includegraphics[width=0.37\textwidth]{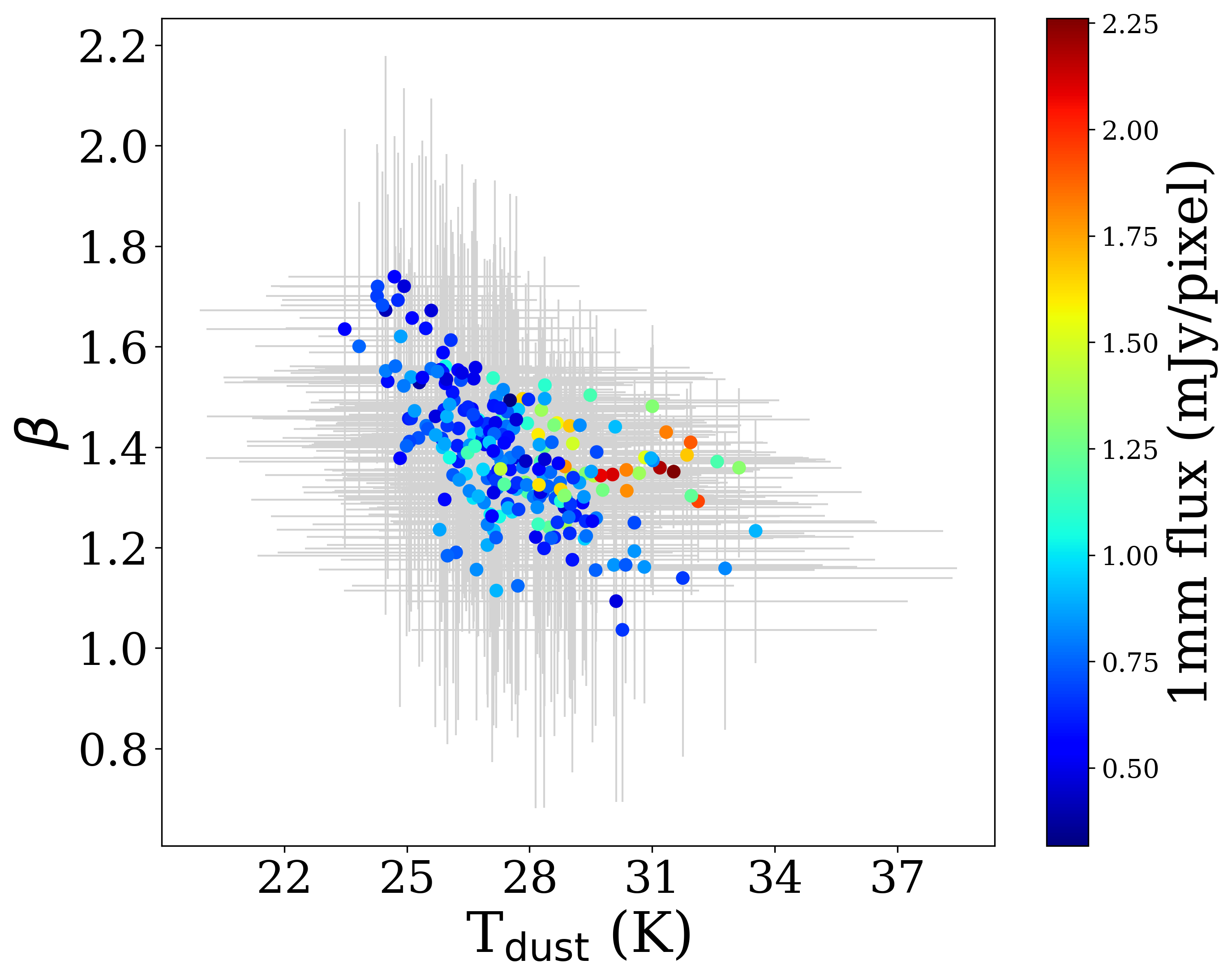}
	\caption{Pixel-by-pixel relation of dust temperature $T_{\rm_{dust}}$ with $\beta$ in (\textit{left}) NGC2146 and (\textit{right}) NGC2976. Only pixels above 3$\sigma$ rms level are included. The color scheme shows the flux density in the observed NIKA2 map at 1.15\,mm.}
	\label{plot-tempbeta}
\end{figure}

We explore the relation between dust emissivity index $\beta$ and temperature $T_{\rm{dust}}$. An anti-correlation between the two has often been reported  (\cite{JuvelaYsard2012},\cite{Paradis+2010}), while \cite{Shetty+2009} cautions it might be created by uncertainties in measurements. Fig.~\ref{plot-tempbeta} shows this relation in both galaxies, including only pixels above 3$\sigma$ level. While the general trend between $T_{\rm{dust}}$ and $\beta$ is an anti-correlation, pixels with relatively higher flux at 1.15\,mm do not completely follow the general inverse trend. In other words, the higher S/N pixels are less affected by the degeneracy between $\beta$ and $T_{\rm{dust}}$ in the MBB model.\\

\begin{figure}
	\centering
        \sidecaption
        \includegraphics[width=0.4\textwidth]{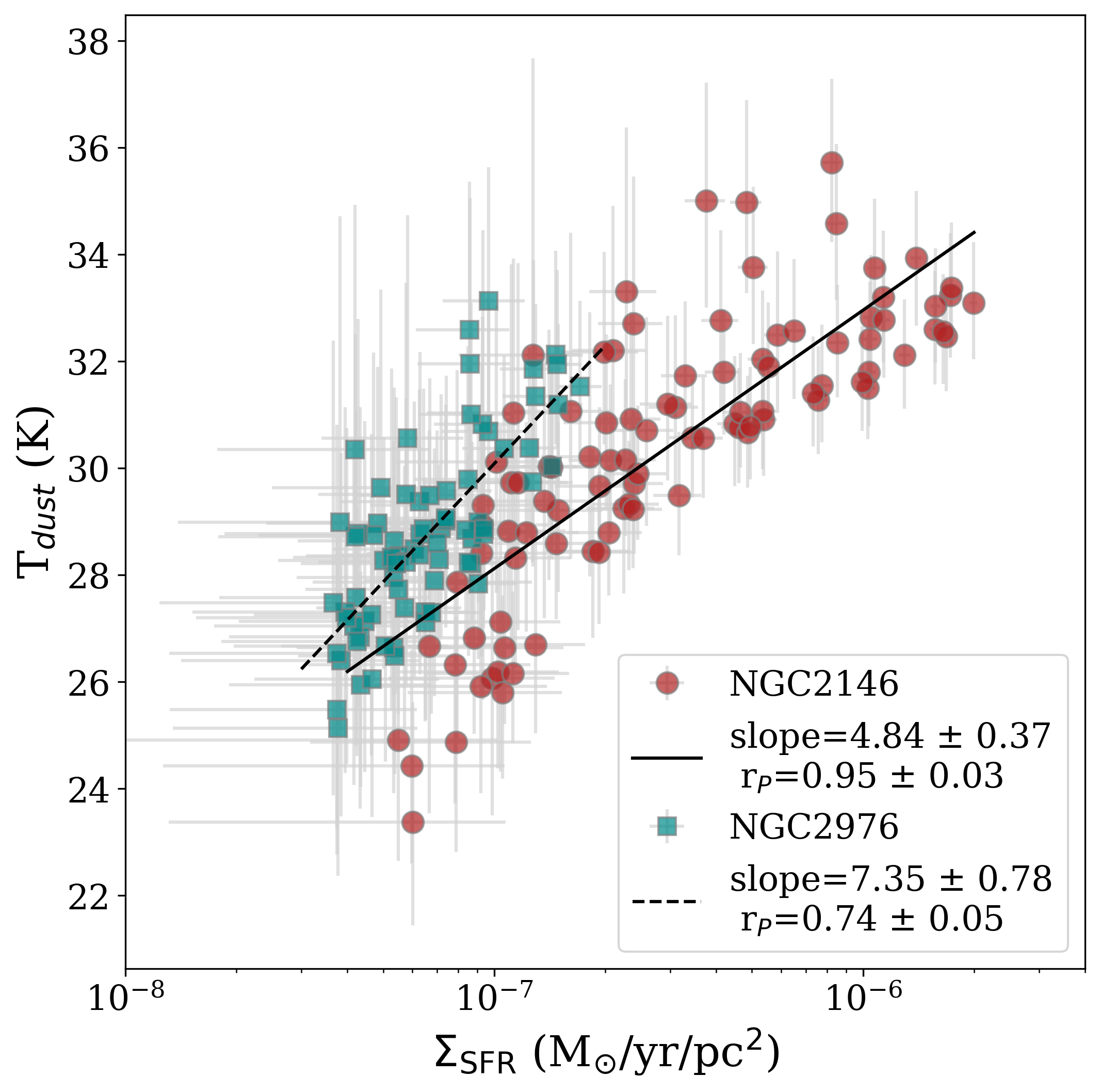}
	\caption{Pixel-by-pixel relation of temperature $T_{\rm{dust}}$ with $\Sigma _{\rm{SFR}}$ in NGC2146 (green dots) and NGC2976 (red dots) for pixels above 5$\sigma$ rms level. The slope of the best-fit line and Pearson coefficient $r_{P}$ are included in the legend.}
	\label{plot-tempSFR}
\end{figure}

We next explore the relation of $T_{\rm{dust}}$ with star formation activity in these galaxies. To do so, we use 24\,$\mu$m observation as a standard tracer of star formation rate density $\Sigma _{\rm{SFR}}$ (\cite{Calzetti+2007}). The relation between $T_{\rm{dust}}$ and $\Sigma _{\rm{SFR}}$ is shown in Fig.~\ref{plot-tempSFR}. The high values found for Pearson correlation coefficients $r_P=0.7$ and $0.9$ indicate tight correlations in both galaxies, even with different ranges of $\Sigma _{\rm{SFR}}$. The positive relation indicates that in regions with higher star-forming activity, InterStellar Radiation Field (ISRF) is intensified by larger amounts of energetic UV photons, heating the dust. Comparison of the slopes in the $T_{\rm{dust}}-\log \Sigma_{\rm{SFR}}$ plane shows that dust temperature increases faster with $\log \Sigma _{\rm{SFR}}$ in the dwarf galaxy NGC2976 than in NGC2146 by a factor of two. This can be linked to their different ISM densities: dwarf galaxies have a less-dense ISM due to lower dust opacity, resulting in a stronger ISRF and faster heating rate by the radiation field. On the contrary, a thicker ISM in a starburst spiral galaxy like NGC2146 means more efficient shielding of dust grains from an energetic radiation field and a slower heating rate \citep{Remy+2015}.

\section{Conclusions}
This research introduces the millimeter observations of two nearby galaxies, NGC2146 (a starburst spiral) and NGC2976 (a dwarf), at 1.15\,mm and 2\,mm, in the framework of the Guaranteed Time Large Project IMEGIN. These observations offer detailed information about the physical characteristics of dust in nearby galaxies by constraining their SED in the FIR-radio wavelength domain. We generate dust mass, temperature, and emissivity index maps of the galaxies and investigate relationships between dust properties and other ISM components such as SFR. An anti-correlation is reported between dust emissivity index and temperature, but the high S/N pixels do not follow the general inverse trend. Additionally, we report a strong correlation between dust temperature and SFR in both galaxies. The impact of star formation activity on dust temperature is more pronounced in the dwarf galaxy NGC2976 than in NGC2146. 

\section*{Acknowledgements}
{\footnotesize 
We would like to thank the IRAM staff for their support during the observation campaigns. The NIKA2 dilution cryostat has been designed and built at the Institut N\'eel. In particular, we acknowledge the crucial contribution of the Cryogenics Group, and in particular Gregory Garde, Henri Rodenas, Jean-Paul Leggeri, Philippe Camus. This work has been partially funded by the Foundation Nanoscience Grenoble and the LabEx FOCUS ANR-11-LABX-0013. This work is supported by the French National Research Agency under the contracts "MKIDS", "NIKA" and ANR-15-CE31-0017 and in the framework of the "Investissements d’avenir” program (ANR-15-IDEX-02). This work is supported by the Programme National Physique et Chimie du Milieu Interstellaire (PCMI) and the Programme National Cosmology et Galaxies (PNCG) of the CNRS/INSU with INC/INP co-funded by CEA and CNES. This work has benefited from the support of the European Research Council Advanced Grant ORISTARS under the European Union's Seventh Framework Programme (Grant Agreement no. 291294). A. R. acknowledges financial support from the Italian Ministry of University and Research - Project Proposal CIR01$\_00010$. S. K. acknowledges support provided by the Hellenic Foundation for Research and Innovation (HFRI) under the 3rd Call for HFRI PhD Fellowships (Fellowship Number: 5357). M.B., A.N., and S.C.M. acknowledge support from the Flemish Fund for Scientific Research (FWO-Vlaanderen, research project G0C4723N.}

\end{document}